\documentclass[12pt]{article}
\newcommand{\bea}{\begin{eqnarray}}
\newcommand{\eea}{\end{eqnarray}}
\begin{document}
\setlength{\baselineskip}{0.7cm}
%

\begin{titlepage}
\begin{flushright}
RIKEN-TH-52~
KEK-TH-1034~
hep-th/0508113
\end{flushright}
\vspace*{5mm}
\begin{center}{\Large\bf Gravitational Radius Stabilization \\
\vspace*{3mm}
in Supersymmetric Warped Compactification}
\end{center}
\vspace{5mm}
\begin{center}
{\large Nobuhito Maru}$^{(a)}$
\footnote{E-mail: maru@riken.jp} and 
{\large Nobuchika Okada}$^{(b)}$
\footnote{E-mail: okadan@post.kek.jp}
\end{center}
\vspace{0.2cm}
\begin{center}
${}^{(a)}$ {\it Theoretical Physics Laboratory, RIKEN, \\
2-1 Hirosawa, Wako, Saitama 351-0198, Japan}
\\[0.2cm]
${}^{(b)}$ {\it Theory Division, KEK, 
Oho 1-1, Tsukuba, Ibaraki 305-0801, Japan \\
Department of Particle and Nuclear Physics, 
The Graduate University \\ 
for Advanced Studies (Sokendai), 
Oho 1-1, Tsukuba, Ibaraki 305-0801, Japan}
\end{center}
\vspace{1cm}
\begin{abstract}
We examine the possibility of the extra dimensional 
radius stabilization with only the gravity multiplet in the bulk 
and some couplings at orbifold fixed points 
in a supersymmetric Randall-Sundrum model. 
Unfortunately, we find that the radius cannot be stabilized 
in all the cases we consider. 
Depending on parameters in the model, 
the fifth dimension collapses or its radius goes to infinity. 
While the former case is theoretically disastrous, 
the latter implies that the so-called ``RS II'' model 
is automatically realized in our setup. 
Although the radius is not stabilized, 
there is nothing wrong with the resultant RS II model, 
because it is not only phenomenologically viable 
but also free from the gauge hierarchy problem 
thanks to its supersymmetric extension. 

\end{abstract}
\end{titlepage}

Brane world scenario is well motivated 
to solve the gauge hierarchy problem without supersymmetry (SUSY). 
Remarkable examples are models with large (flat) extra dimensions 
\cite{ADD, CIM,Kolorelis} 
and the model with a warped extra dimension \cite{RS1}. 
Alternatively, an extension to SUSY brane world scenario is also well 
motivated to solve SUSY flavor problem geometrically \cite{RS3, LS1}. 
Irrespective of the existence of SUSY or not, 
the radius stabilization is a generic important issue 
to construct realistic brane world scenarios. 
So far, many models have been proposed in non-SUSY case 
\cite{GW, GPT, BMNO, PP, SS, GP, GH} and SUSY case 
\cite{LS1, LS2, LO, GQR, MO, EMS, KY, Blechman, DQ, GH}. 
All these models introduce new bulk fields in addition to gravity, 
which play the crucial role to succeed the radius stabilization. 
Introduction of such bulk fields makes the model complicated 
and somewhat artificial. 
It would be nice if the radius can be stabilized 
via only the gravitational effects with 
only gravity in the bulk and some couplings 
among the radius moduli (radion) and the brane fields. 
This possibility through the Casimir energy was examined 
many years ago and found to be failed \cite{AC}.  
In models with flat extra dimensions, 
the existence of the brane does not make the situation better 
since there is no coupling among the radion and brane fields 
at tree level. 
On the other hand, in the model with warped extra dimension, 
the brane fields can couple to the radion at tree level 
and such couplings may play an important role 
for the radius stabilization with only gravitational effects. 
In this paper, we examine this possibility 
in a SUSY Randall-Sundrum (RS) model \cite{ABN, GhP, FLP, BKV}.

Let us consider the SUSY RS model with the ${\rm AdS}_5$ metric 
\bea
ds^2 = e^{-2kry}\eta_{\mu\nu}dx^\mu dx^\nu 
- r^2dy^2~(\mu, \nu =0,1,2,3), 
 \label{metric} 
\eea
where $k,x,y,r$ are an ${\rm AdS}_5$ curvature scale, 
the coordinates of four dimensions, 
the angle and the radius of the fifth dimension, $0 \le y \le \pi$, 
respectively. 
Here the ${\rm AdS}_5$ curvature scale should be small 
compared to the five dimensional Planck scale $M_5$,
so that the above metric can be trusted. 
In our setup, the bulk field is only the gravity multiplet 
and the other fields reside only on branes at $y=0,\pi$ 
being $S^1/Z_2$ orbifold fixed points. 

As the simplest setup to arise the nontrivial radion potential, 
let us first examine the case with constant superpotentials 
on the both boundary branes. 
In the superconformal framework, the four dimensional effective Lagrangian 
below the Kaluza-Klein (KK) scale is given by \cite{MP}
\bea
{\cal L} &=& \int d^4 \theta 
\left[ -\frac{3M_5^3}{k} (\phi^\dag \phi - \omega^\dag \omega) \right] 
+ \left[ \int d^2 \theta (\phi^3 W_0 + \omega^3 W_\pi) + {\rm h.c.} \right], 
\eea
where $\phi$ denotes the compensating multiplet $\phi = 1 + \theta^2 F_\phi$, 
$\omega \equiv \phi e^{-k \pi T}$ ($T:$ the radion multiplet, ${\rm Re}T=r$), 
and $W_{0,\pi}$ are the constant superpotentials localized at $y=0, \pi$. 
The Lagrangian for auxiliary fields can be read off as
\bea
{\cal L}_{{\rm aux}} &=& 
F_\phi^\dag \left\{ -\frac{3M_5^3}{k} F_\phi + 3W_0^\dag \right\}
+ F_\omega^\dag \left\{ \frac{3M_5^3}{k} F_\omega 
+ 3(\omega^2 W_\pi)^\dag \right\} \nonumber \\
&&+ 3F_\phi W_0 + 3 F_\omega \omega^2 W_\pi. 
\label{aux1}
\eea
Equations of motion for auxiliary fields give solutions
\bea
F_\phi = \frac{k}{M_5^3}W_0^\dag, \quad
F_\omega = -\frac{k}{M_5^3} (\omega^2 W_\pi)^\dag. 
\eea
Substituting $F_\phi$ and $F_\omega$ into (\ref{aux1}), 
the radion potential is obtained as
\bea
V &=& -3F_\phi W_0 -3 F_\omega \omega^2 W_\pi,\\
&=& \frac{3k}{M_5^3}\left(- |W_0|^2 + |\omega^2 W_\pi|^2 \right), 
\eea
which implies that the radius goes to infinity $\omega=0$. 
For consistency with the metric, 
namely the flat 4 dimensional spacetime metric, 
the cosmological constant should vanish,  
so that $W_0$ should be zero or a (global) SUSY breaking source 
to cancel the negative cosmological constant should be added. 
Then, the model automatically reduces 
to the so-called ``RS II'' model \cite{RS2}.

The next possibility to stabilize the radius without introducing 
any bulk field is that there exit SUSY breaking fields 
localized at $y=\pi$ brane. 
Note that any brane field at $y=0$ has no effects 
for the radius stabilization since it does not couple to the radion. 
Thus, two parts involving $\phi$ and $\omega$ are completely 
separated in the Lagrangian, 
and the part relevant to the radion potential is 
\bea
{\cal L} = \int d^4\theta 
\left[
\left( \frac{3M_5^3}{k} + K(X^\dag, X) \right) \omega^\dag \omega 
\right]
+ \left[ \int d^2\theta \omega^3 W(X) + {\rm h.c.} \right], 
\eea
where $X$ is a chiral multiplet localized at $y=\pi$ brane, 
their K\"ahler potential and superpotential are denoted 
as $K(X^\dag, X)$ and $W(X)$. 
The auxiliary components of the above Lagrangian can be extracted as
\bea
{\cal L}_{{\rm aux}} 
&=& F_\omega^\dag \left( \left(\frac{3M_5^3}{k} +K \right) F_\omega 
 + K_X F_X \omega + 3 (\omega^2 W)^\dag \right) \nonumber \\
&&+F_X^\dag \left( K_{X^\dag} F_\omega \omega^\dag 
+ K_{XX^\dag} F_X |\omega|^2 + (\omega^3 W_X)^\dag \right) 
+ 3 \omega^2 F_\omega W + \omega^3 W_X F_X, 
\eea
where $K_X$ etc. are the shorthand notation of 
the derivative of $K$ with respect to $X$ etc. 
Equations of motion for the auxiliary fields give solutions, 
\bea
\left(
\begin{array}{c}
F_\omega \\
F_X
\end{array}
\right)
=\frac{1}{|B|^2 - AC} 
\left(
\begin{array}{c}
3C (\omega^2 W)^\dag -B(\omega^3 W_X)^\dag \\
-3B^\dag (\omega^2 W)^\dag + A (\omega^3 W_X)^\dag \\
\end{array}
\right), 
\eea
where 
\bea
A &=& \frac{3M_5^3}{k} + K,\\
B &=& K_X \omega,\\
C &=& K_{XX^\dag} |\omega|^2. 
\eea
Substituting these into ${\cal L}_{{\rm aux}}$, 
we obtain the radion potential 
\bea
\label{Xpot}
V &=& -3 \omega^2 F_\omega W(X) - \omega^3 W_X F_X, \\
&=& \frac{|\omega|^4}{AC-|B|^2} 
\left[
C \left| 3 W(X)^\dag -\frac{B W_X^\dag}{K_{XX^\dag} \omega} \right|^2 
+ C^{-1} \left(AC-|B|^2 \right) |\omega|^2 |W_X|^2
\right]. 
\label{Xpot1}
\eea
The general argument is the following. 
The Lagrangian we consider has the same form as the global SUSY one, 
so that the radion potential is semi-positive definite. 
Noting that the potential is proportional to $\omega$, 
we can conclude that the global minimum of the potential $V=0$
is always realized at $\omega =0$. 
Therefore, the radius goes to infinity ($\omega=0$) and 
the model reduces to the SUSY RS II model again. 
Here we have used the fact that the K\"ahler metric 
should be positive, 
namely $AC-|B|^2 > 0$ and $A, C >0$ 
in the explicit form of the potential (\ref{Xpot1}), 
otherwise the system itself is unstable.
\footnote{Even if $AC-|B|^2 < 0$, one can easily show that 
the vacuum at the finite radius is the local minimum.}

The final possibility we consider is to include quantum effects, 
namely 1-loop Casimir energy 
which has been recently calculated \cite{GRSST} in the different context. 
We expect that the radius can stabilized in this case, 
since there are nontrivial radius dependences 
in the potential at tree and one-loop levels.\footnote{
Even if we find that the Casimir effect works so as to stabilize
the radius, we have to be careful for our treatment
because the effect is the corrections of the next order.
As will be seen later, we come to the same result as in the tree
level when the Casimir effects are taken into account. 
} 
In this case, the Lagrangian becomes complicated 
since the mixing term between $\phi$ and $\omega$ 
in the K\"ahler potential is induced by the 1-loop Casimir effects. 
The Lagrangian we examine is given by
\bea 
{\cal L}= \int d^4 \theta K + 
\left\{  \int d^2 \theta W  + h.c. \right\} , 
\eea 
where 
\bea 
K = -\frac{3 M_5^3}{k} (\phi^\dagger \phi - \omega^\dagger \omega)  
 + \frac{c k^2}{4 \pi^2} 
 \frac{(\omega^\dagger \omega)^2}{\phi^\dagger \phi},~
W = \phi^3 W_0 + \omega^3 W_\pi .  
\eea 
The last term in the K\"ahler potential is the new term 
 induced by 1-loop corrections\footnote{We consider only the simpler form 
 in the strongly warped case, for simplicity.} and $c \simeq 1.165$ 
 is the numerical number \cite{GRSST}. 
The Lagrangian for auxiliary fields can be read off as
\bea 
{\cal L}_{{\rm aux}} &=& 
 F_\phi^\dagger \left( 
  K_{\phi^\dagger \phi}  F_\phi + 
  K_{\phi^\dagger \omega} F_\omega + 3 W_0^\dagger 
  \right)  \nonumber \\ 
 &+&  F_\omega^\dagger  \left( 
  K_{\omega^\dagger \phi}  F_\phi + 
  K_{\omega^\dagger \omega} F_\omega + 3 (\omega^2 W_\pi)^\dagger 
 \right) +3 W_0 F_\phi +3 \omega^2 W_\pi F_\omega, 
\eea 
where
$K_{\phi^\dagger \phi}$, 
$K_{\phi^\dagger \omega}$ etc. 
 are explicitly described as follows: 
\bea 
 K_{\phi^\dagger \phi} &=& \left. 
 \frac{\partial^2 K}{ \partial \phi^\dagger \partial \phi} 
\right|_{\phi=1} 
= - \frac{3 M_5^3}{k} + \frac{c k^2}{4 \pi^2} (\omega^\dagger \omega)^2 , 
\nonumber \\ 
 K_{\phi^\dagger \omega} &=&  
 \left(  K_{\omega^\dagger \phi} \right)^\dagger =  
\left. 
 \frac{\partial^2 K}{ \partial \phi^\dagger \partial \omega} 
\right|_{\phi=1} 
= - \frac{c k^2}{2 \pi^2} (\omega^\dagger \omega) \omega^\dagger ,
\nonumber \\ 
 K_{\omega^\dagger \omega} &=& 
\left. 
 \frac{\partial^2 K}{ \partial \omega^\dagger \partial \omega} 
\right|_{\phi=1} 
= \frac{3 M_5^3}{k} + \frac{c k^2}{\pi^2} (\omega^\dagger \omega) . 
\eea  
Equations of motion for auxiliary fields give solutions, 
\bea
F_\phi &=& \frac{k}{M_5^3} 
 \left(
 1 + \tilde{c} (\omega^\dagger \omega) 
   - \frac{\tilde{c}}{4} (\omega^\dagger \omega)^2 
 \right)^{-1}  
 \left[ 
  \left( 1+\tilde{c} (\omega^\dagger \omega)  \right) W_0^\dagger 
 + \frac{\tilde{c}}{2} 
  (\omega^\dagger \omega) ( \omega^3 W_\pi)^\dagger 
\right] ,  \\ 
F_\omega &=& - \frac{k}{M_5^3} 
 \left(
 1 + \tilde{c} (\omega^\dagger \omega) 
   - \frac{\tilde{c}}{4} (\omega^\dagger \omega)^2 
 \right)^{-1}  
  \left[ 
  -  \frac{\tilde{c}}{2} 
   (\omega^\dagger \omega) \omega W_0^\dagger 
  + \left(  1 - \frac{\tilde{c}}{4} (\omega^\dagger \omega)^2  \right) 
 ( \omega^2 W_\pi)^\dagger  
\right]   , 
\eea 
where $\tilde{c} = \frac{c k^3}{3 \pi^2 M_5^3} 
 \leq \frac{c}{3 \pi^2} \ll 1$ for $ k \leq M_5$. 
Substituting these into the radion potential, we obtain 
\bea 
 V &=& -3 W_0 F_\phi -3 \omega^2 W_\pi F_\omega, \\
 &=& \frac{3 k}{M_5^3}   \left( 
  1 + \tilde{c} (\omega^\dagger \omega) 
   - \frac{\tilde{c}}{4} (\omega^\dagger \omega)^2  \right)^{-1} 
 \nonumber \\ 
 & \times &  
\left[
 - \left( 1+ \tilde{c} (\omega^\dagger \omega)  \right) 
   \left| W_0 \right|^2 
 + \left( 1 - \frac{\tilde{c}}{4} (\omega^\dagger \omega)^2   \right) 
  (\omega^\dagger \omega)^2  \left| W_\pi \right|^2 
  \right.  \nonumber  \\ 
 &-& \frac{\tilde{c}}{2} (\omega^\dagger \omega) \left. 
 \left( 
 W_0 (\omega^3 W_\pi)^\dagger + W_0^\dagger (\omega^3 W_\pi) 
\right) 
\right] .
\label{potential}
\eea

Before analyzing the radion potential in detail, 
let us examine the radius stabilization for two special cases. 
The first is the case with $W_\pi=0$. 
The radion potential is reduced to the form,  
\bea 
 V = -  \frac{3 k}{M_5^3}   \left( 
  1 + \tilde{c} (\omega^\dagger \omega) 
   - \frac{\tilde{c}}{4} (\omega^\dagger \omega)^2  \right)^{-1} 
  \left( 1+ \tilde{c} (\omega^\dagger \omega)  \right) 
   \left| W_0 \right|^2. 
\eea 
Since $\tilde{c} \ll 1$ and  $\omega$ is defined 
in the range $ 0 \leq \omega \leq 1$, 
the potential has a minimum at $\omega =1$. 
This means that the fifth dimension collapses. 
On the other hand, in the case with $W_0=0$, 
the radion potential is found to be 
\bea 
 V =  \frac{3 k}{M_5^3}   \left( 
  1 + \tilde{c} (\omega^\dagger \omega) 
   - \frac{\tilde{c}}{4} (\omega^\dagger \omega)^2  \right)^{-1} 
 \left( 1 - \frac{\tilde{c}}{4} (\omega^\dagger \omega)^2   \right) 
  (\omega^\dagger \omega)^2  \left| W_\pi \right|^2 . 
\eea  
The potential minimum is realized at $\omega=0$ 
 and the potential energy is zero there, 
 which means the fifth dimensional radius 
 goes to infinity and the system becomes the SUSY RS II model. 
Since the above two cases come to opposite results, 
 we may expect that the radius can be stabilized at $0 < \omega < 1$ 
 if we introduce both non-zero $W_0$ and $W_\pi$.

Now let us examine the case with both non-zero constant superpotentials. 
Considering the form of the potential (\ref{potential}), 
 we can start with both $W_{0,\pi}$ real and positive 
 without loss of generality 
 by an appropriate convention of the complex phase of $\omega$. 
Parameterizing $\omega=x e^{i \theta/3}$ ($0 \leq x=|\omega| \leq 1$), 
 $\theta$-dependent part in the potential is given by 
\bea
 V \supset - \frac{3 k}{M_5^3}   \left( 
  1 + \tilde{c} x^2 - \frac{\tilde{c}}{4} x^4 \right)^{-1} 
  \left( \tilde{c} W_0 W_\pi x^5 \cos \theta 
  \right) . 
\eea
Thus, the potential minimum in $\theta$-direction is realized at $\theta=0$. 
For examining the potential minimum in $x$-direction, 
 it is useful to rewrite the potential into the form, 
\bea
\hat{V}(x) = \frac{V}{ \frac{3 k}{M_5^3} W_0^2 } 
= f(x)^{-1} g(x) , 
\eea 
where 
\bea 
 f(x) &=& 1 + \tilde{c} x^2  - \frac{\tilde{c}}{4} x^4 , \nonumber \\
 g(x) &=&  - \left( 1+ \tilde{c} x^2  \right) 
   + \left( 1 - \frac{\tilde{c}}{4} x^4   \right) x^4 a^2 
   - \tilde{c} x^5 a 
\eea
with $a= W_\pi/W_0 >0$. 
Since $f(x) > 0$ for $\tilde{c} \ll 1$ and $0 \leq x \leq 1$, 
 the minimization condition is equivalent to the condition $f g_x -f_x g =0$, 
 where $f_x$ etc. stand for $df/dx$ etc. 
For $0< x <1$, the condition is explicitly described as 
\bea 
 \frac{f g_x-f_x g}{x^3} 
 &=& - \left( \tilde{c} + \frac{\tilde{c}^2}{2} x^2 \right)  
     + \left( -5 \tilde{c} x -3 \tilde{c}^2 x^3 + \frac{\tilde{c}^2}{4}x^5 
       \right) a  \nonumber \\
 &+& \left( 
   4+ 2 \tilde{c} x^2 -2 \tilde{c} x^4 -
   \frac{3 \tilde{c}^2}{2} x^6 + \frac{\tilde{c}^2}{4} x^8
    \right) a^2   \nonumber \\ 
 &\simeq&  - \tilde{c} -5 \tilde{c} a x +4 a^2 =0,  
\eea
 where we have used $\tilde{c} \ll 1$ and $ 0< x < 1$ 
 in the approximation formula. 
The minimization condition is satisfied by 
 $x \simeq x_0 = (4 a^2-\tilde{c})/(5 \tilde{c} a)$. 
Since the second derivative of the potential at $x=x_0$ 
 is given by 
 $\hat{V}_{xx}|_{x=x_0} =f^{-2} (f g_{xx}-f_{xx}g)|_{x=x_0}$,
 the sign of $V_{xx}$ at $x=x_0$ is determined by 
 $(f g_{xx}-f_{xx}g)|_{x=x_0}$. 
Its explicit form is described as 
\bea
 \frac{f g_{xx}-f_{xx}g |_{x=x_0}}{x_0^2}
 &=& - \left(3 \tilde{c} + \frac{5 \tilde{c}^2}{2} x_0^2 \right) 
  + \left( -20 \tilde{c} x_0 - 18 \tilde{c}^2 x_0^3 + 2 \tilde{c}^2 x_0^5 
    \right) a  \nonumber \\
 &+& \left( 
   12+ 10 \tilde{c} x_0^2 -14 \tilde{c} x_0^4 
   - \frac{27 \tilde{c}^2}{2} x_0^6 + \frac{11 \tilde{c}^2}{4} x_0^8
    \right) a^2   \nonumber \\ 
 &\simeq& - 3 \tilde{c} -20 \tilde{c} a x_0 + 12 a^2 
 = -5 \tilde{c} a x_0 < 0  .
\eea 
Therefore, the potential has its maximum at $x=x_0$ 
 and the radius is not stabilized. 
The global potential minimum is realized at $x=0$ or $x=1$ 
 depending on the value $a$. 
By straightforward calculations, we can find the global minimum 
 at $x=0$ ($x=1$) for $ a \geq a_c$ ($a \leq a_c$), 
 where $a_c= \frac{2 \tilde{c} +\sqrt{3 \tilde{c}^2 +4 \tilde{c}}}
 {4-\tilde{c}}$.
Even in either case, since the vacuum energy is found to be negative, 
a global SUSY breaking source should be added to obtain 
the vanishing cosmological constant.

In summary, 
we have investigated whether the extra dimensional radius can be stabilized 
via only the gravitational effects in the SUSY Randall-Sundrum model. 
We have found that the radius cannot be stabilized 
even if we take into account 
the constant superpotentials on the branes, 
the couplings with the brane localized SUSY breaking field, 
and the 1-loop Casimir energy effects. 
Depending on the parameters in the model, 
the fifth dimension collapses or its radius goes to infinity. 
While the former case is theoretically disastrous, 
the latter implies that the RS II model can be 
automatically realized in our setup. 
Although the radius cannot be stabilized, 
there is nothing wrong with the resultant SUSY RS II model, 
because it is not only phenomenologically viable 
and also free from the gauge hierarchy problem 
thanks to its SUSY extension. 
Moreover, the RS II model has an interesting aspect on cosmology. 
Its cosmological solution have been found to lead to 
a non-standard Friedmann equation, and the expansion rate 
can be altered in the early universe \cite{RSIICosmo}. 
Some particle cosmological implications of the RS II cosmology 
have been discussed elsewhere \cite{OS, NOS}. 

Finally, we give a comment on more complicated cases. 
Considering the Casimir effects (including higher order corrections) 
in the case with both boundary fields and 
their non-trivial K\"ahler potentials and superpotentials 
is the most general setup to examine the radius stabilization. 
Since, in this case, there are many mixing terms 
among all the fields including $\phi$ \cite{GRSST}, 
it is very hard to analyze this case and get an definite conclusion. 
We have skipped this case in the paper. 
However, note that, if $ k \ll M_5$, 
Casimir effects are negligible and 
we come to one of our conclusion in the paper.

\vskip 1cm
\begin{center}
{\bf Acknowledgments}
\end{center}  
The work of N.M. is supported by RIKEN (No. A12-61014). 
The work of N.O. is supported in part by the Grant-in-Aid for Scientific 
Research in Japan (No.15740164). 
\vskip 1cm

\end{document}